\documentclass[12pt,a4paper]{article}

\usepackage{amsmath}
\usepackage{graphicx}
\usepackage{times}
\usepackage{cite}
\usepackage{ulem}
\begin{document}
\begin{titlepage}
\title{{ Centrality  in  hadron collisions}}
\author{ S.M. Troshin, N.E. Tyurin\\[1ex]
\small  \it NRC ``Kurchatov Institute''--IHEP\\
\small  \it Protvino, 142281, Russian Federation.}

\normalsize
\date{}
\maketitle

\begin{abstract}
In hadron interactions at the LHC energies,  the reflective scattering mode  starts to play a noticeable  role which is expected to be even a more significant beyond the energies of the LHC. This new but still arguable phenomenon implies   a peripheral dependence of the inelastic probability distribution in the impact parameter space and asymptotically evolving to the black ring.  As a consequence, the straightforward extension to hadrons of
the  centrality definition  adopted for nuclei  needs to be modified. 
\end{abstract}
\end{titlepage}
\setcounter{page}{2}

Hadrons are  the extended objects, i.e. their formfactors  are described by nontrivial functions. Despite the hadrons have a smaller size compared to the nuclei, one could expect similarity in hadrons' and nuclei's interactions, and observation
of  collective phenomena in both  cases is in favor of this similarity. Of course, such a likeness   occurs at the different energies in small and large systems\footnote{Results of experimental studies  of nuclear interactions at RHIC and hadron interactions at the LHC are in favor of such conclusion} despite that   hadronic matter distribution has a similar form in hadrons and nuclei. Here the energy--dependent interaction dynamics plays a significant role. Interest in the collective effects studies in both large and small systems is justified by  their relation to  the dynamics of  quark-gluon plasma formation \cite{rafel}. In addition, confinement of color in hadrons is also associated with collective, coherent interactions of quarks and gluons. 

The hadron scattering amplitude as well as  the elastic and inelastic overlap functions gets contributions  associated with  geometry and dynamics of a collision. 
The LHC  experiments have led to  discovery of collective effects  in  small systems \cite{cms} (for  comprehensive list of references to the experimental results of ALICE, ATLAS, CMS and LHCb Collaborations see \cite{mangano} and for a brief review --- \cite{trib}).  A well  known example is an observation of a "ridge" effect in two--particle  correlations in 
 $pp$-collisions in the events with high multiplicities \cite{weili, weili1}. To reveal the dynamics of a collective state formation one needs to answer the question: what are the impact parameter values responsible for such events? To answer it one needs to be able to reconstruct these values from the experimental observables since the impact parameter is  not measurable directly. Thus, definition of centrality representing the collision geometry becomes a crucial issue in small systems studies also.

 The reflective scattering mode leads to  formation of a peripheral impact parameter dependence of the inelastic interactions probability because of unitarity. This mode  significantly affects collisions with small impact parameters suppressing production of secondary particles in such collisions and hence it is important  for assignment of  a correct value of the impact parameter for  particular hadron collision events.  
 
 Emergent ring-like dependence of the inelastic overlap function can be associated with the effect of self--dumping intermediate inelastic channels contributions at the LHC energies. Such self-dumping can arise as a result of a randomization of the phases of  multiparticle production amplitudes.  This randomization in its turn can be considered as a consequence of the color--conducting collective state of hadronic matter formation under the central over impact parameter hadron collisions  (with high multiplicities) since  a  subsequent stochastic decay of this state and hadronization  into the multiparticle final states  takes place.  This is just the energy range where reflective scattering is most significant and where peripheral form of the inelastic overlap function should be accounted.

 Here we discuss how to improve definition of centrality  for small systems (pp-interactions) if appropriate and the role of the reflective mode  at the LHC energy range and beyond. 
 
  The centrality  is  a commonly accepted variable for description and classification of the collision events in the nuclei interactions. This variable is related to the assumed initial--state collision geometry and is given by an impact parameter value associated with  the general geometrical characteristics of a particular  collision event.We suppose that meaning of centrality for hadron collisions can be used in the same way as it is applied in the case of nuclei-- nuclei and hadron--nuclei collisions.
 
  In the experiments with nuclei (including hadron-nucleus reactions)  (see\cite{olli} and \cite{olli1}), the   centrality $c^A_b$ is extracted from either number of charged particles registered in the respective detector or the transverse energy deposed into the calorimeter. Those quantities  ( both denoted by $n$) are relevant to  experimentally measurable quantity   $c^A$ also called centrality.   Superscript $A$ means nuclear collisions while superscript $h$ denotes pure hadron collisions. The definitions of $c^A_b$   \cite{olli} is
 \begin{equation}
 \label{centb}
 c_b^A\equiv \frac{\sigma^b_{inel}}{\sigma_{inel}},
 \end{equation}
 where 
 \[
 \sigma^b_{inel}=\int_0^bP^A_{inel}(b')2\pi b'db
 \]
 and  $P^A_{inel}(b)$ is probability distribution of the inelastic collisions  over the impact parameter $b$, whereas  the experimentally measurable quantity 
 \begin{equation}
 \label{cent}
 c^A(n)\equiv \int_n^\infty P^A(n')dn',
 \end{equation}
 where $P_A(n')$ is the probability for the quantity $n$ have the value $n'$.
 is based on distribution over the multiplicity or the total transverse energy of the final state \cite{olli}. 

 It should be noted that the energy dependence of the above quantities is tacitly implied and not indicated explicitly. The energy dependence, however, can be a nontrivial one in the collisions of nuclei as well as of hadrons, since size of interaction region, probabilities of interactions, multiplicities and transverse energies are the energy--dependent quantities in both cases. Evidently, the effects related to the energy dependence of all these quantities should be taken into account under analysis of the experimental data with the same values of centrality but measured at the different energies.
 
In view of the prominent collective effects observed in small systems, such as $pp$-collisions, together with indication on the reflective scattering mode presence  at the LHC, an introduction of centrality notion  accounting these phenomena  is necessary  to classify the  events of interest. 
 
 The hadron scattering has  similarities as well as differences with the scattering of nuclei.  Geometrically, hadrons are the extended objects too, but a rather significant contribution to $pp$--interactions is provided by the elastic scattering with the ratio of elastic to total cross-sections $\sigma_{el}(s)/\sigma_{tot}(s)$  increasing with energy. The elastic scattering of nuclei is not so important at high energies since nucleons are not confined in nuclei.  Hence, the geometrical characteristics of hadron collisions related to the elastic scattering   are essential   for  the hadron dynamics understanding, i.e. for the  development of QCD in the soft region where the confinement  and collective effects play a crucial role. 
 
 It will be argued further that the  definition of centrality based on the use of a straightforward analogy with nuclei interactions is not suitable for the hadrons at the energies where the reflective scattering gives a significant contribution. To obtain a relevant definition, we  propose to use a full probability distribution $P^h_{tot}(s,b)$ in order to take into account the elastic  events. The neglect of the elastic  events would lead to incorrect estimation of centrality for the  hadron collision.
 For the centrality $c_b^h(s,b)$ the following definition   is suggested
 \begin{equation}
 \label{centhb}
 c_b^h(s,b)\equiv \frac{\sigma^b_{tot}(s)}{\sigma_{tot}(s)},
 \end{equation}
 where
 \[
 \sigma^b_{tot}(s)=8\pi\int_0^b\mbox{Im}f(s,b') b'db'
 \]
 is the impact--parameter dependent cumulative contribution into the total cross--section, $\sigma^b_{tot}(s)\to \sigma_{tot}(s)$ at $b\to\infty$.  In Eq. (\ref{centhb}) the total (elastic plus inelastic) contribution replacing the inelastic cross--section only were used.
 
  It should be noted, that there is no problems with definition of centrality in the form of
 Eq. (\ref{centb}) in case of hadron scattering at the energies where the reflecting scattering mode is not present,
 but its possible presence  at higher energies would change form of an inelastic overlap function $h_{inel}(s,b)$ from a central to  peripheral one with maximum at $b\neq 0$. Therefore,  Eq. (\ref{centb}) for centrality in this case needs to be modified. The use of centrality in this form  will not correctly represent an impact parameter  dependence and  reflect  collision geometry. 
 
 Contrary to the inelastic overlap function $h_{inel}(s,b)$ , the function $\mbox{Im}f(s,b)$ always and  at the LHC energies, in particular,  has a central impact parameter profile with a maximum  located at $b=0$ \cite{alkin1}. The paper \cite{alkin1} indicates that $\mbox{Im}f(s,b)>1/2$ at $b=0$.
 The amplitude $f(s,b)$ is the Fourier--Bessel transform of the scattering amplitude $F(s,t)$:
 \begin{equation}\label{imp}
 F(s,t)=\frac{s}{\pi^2}\int_0^\infty bdbf(s,b)J_0(b\sqrt{-t}).
 \end{equation}
 The definition  Eq. (\ref{centhb}) can be inverted, namely,
 one can consider centrality as an observable measured  in hadron collisions. Eq. (\ref{centhb}) then can be used for  restoration of  the elastic scattering amplitude. More specifically, the function $\mbox{Im} f(s,b)$ can be calculated, if the impact parameter dependence of  $c_b^h(s,b)$ is experimentally known.
 The inverted relation corresponding to Eq. (\ref{centhb}) written in the differential form gives:
 \begin{equation}
 \label{centhbder}
 \mbox{Im} f(s,b)=\frac{\sigma_{tot}(s)}{8\pi b}\frac{\partial c_b^h(s,b)}{\partial b} 
 \end{equation}
 and, as it was said above, could be instrumental for the reconstruction of $\mbox{Im} f(s,b)$ from the inelastic collisions. However, a practical feasibility of such reconstruction needs further studies.

 The impact parameter representation provides a simple semiclassical picture of hadron scattering, e.g. head--on or central collisions correspond to small impact parameter values.
 From Eq. (\ref{centhbder}) one can easily get the inequality
 \begin{equation}
 0\leq\frac{\partial c_b^h(s,b)}{\partial b}\leq \frac{8\pi b}{\sigma_{tot}(s)}
 \end{equation}
 or in the integral form
 \begin{equation}
 0\leq { c_b^h(s,b)}\leq \frac{4\pi b^2}{\sigma_{tot}(s)}
 \end{equation}
 for $b\leq r(s)$, $r(s)\sim {1}/{\mu} \ln s$, where $S(s,b=r(s))=0$, $s>s_r$ and $\mu$ is determined by    mass of pion.

  Reflective scattering mode appears in  the unitarization scheme representing scattering amplitude $f(s,b)$ in the rational form (one-to-one transform). It allows amplitude variation in the whole  region allowed by unitarity \cite{umat}.
 Respective form of the function $S(s,b)$ is written as a Cayley transform\footnote{This one-to-one transform maps upper 
 	half-- plane into a unit circle in case when $U$ and $S$ both are complex functions and the value of the function $S=0$ is reached at finite values of energy and impact parameter.} :
 \begin{equation}
 S(s,b)=\frac{1-U(s,b)}{1+U(s,b)}. \label{umi}
 \end{equation}
  The real, nonnegative function $U(s,b)$ can be considered as an  input amplitude which is subject to further unitarization procedure. This function gets contribution from all inelastic channels, elastic channel does not contribute into the function $U$. The multi-Reggeon processes where small groups of particles are separated
  by large rapidity gaps discussed in \cite{rys} also contribute to the the function $U$. Using connection of the direct and annihilation channels of reaction by means of $U$, i.e. $U\to \tilde U$, discussed in (cf. \cite{umat} and references therein) allows one, in principle, to solve the problem with unitarity\footnote{The issue is relevant also to the energy dependence of the gap survival probability. Modification of the gap survival probability definition for the case of reflective scattering has been proposed in \cite{2008}} described in \cite{rys}. Moreover, we adhere here to the point of view described in \cite{gas} that the power-like contribution (which  is here a sum of terms each of them is proportional to $(\ln s)^n$, where $n$ here stands for a number of particles $a_i$ in the process:
  $$
 p+p\to p+a_1+a_2+...+a_n+p)
   $$ of such exclusive processes should be included into the potential or the function $U$ and, therefore, not be considered as a violation of the Froissart--Martin bound or unitarity. Instead, such  rising fast with energy contribution emphasizes the necessity for the utilization of the  unitarization scheme for obtaining the elastic scattering amplitude. What has been said above is valid for the amplitudes of exclusive production processes. Of course, the amplitudes of such processes need separate treatment, but the situation as a whole is similar to the one  when Pomeron has intercept greater than unity and therefore unitarization is needed.
  
  Thus, the various models provide a monotonically increasing energy
 dependence of the function $U(s,b)$  (e.g. power-like one) and its exponential decrease with  the impact parameter (due to analyticity in the Lehmann-Martin ellipse).
 The value of the  energy corresponding to a complete absorption of the initial state
 at the central collisions  $S(s,b)|_{b=0}=0$
 is denoted as $s_r$ and  determined by the  equation
 $U(s_r,b)|_{b=0}=1$\footnote{The old (pre--LHC) numerical estimates  of $s_r$ have given for its value  $\sqrt{s_r}=2-3$ $TeV$ \cite{srvalue}.}. Beyond this energy the function  $S(s,b)|_{b=0}$ becomes negative and reflective scattering mode appears.

Finally, we discuss the energy dependence of centrality.
 A wide class of the geometrical models for hadron interactions (relevant for the centrality discussion in these processes) allows one to assume that $U(s,b)$ has a factorized form (cf. \cite{factor} and references therein):
 \begin{equation}\label{usb}
 U(s,b)=g(s)\omega(b),
 \end{equation}
 where $g(s)\sim s^\lambda$ at the large values of $s$, and the power dependence guarantees asymptotic growth of the total cross--section $\sigma_{tot}\sim \ln^2 s$. Such a factorized form corresponds to a common source for the increase with energy of the total cross--sections and the slope of the diffraction cone in  elastic scattering due to unitarization \cite{factor}. The particular  form of the function $\omega(b)\sim \exp(-\mu b)$ is consistent with analytical properties of the scattering amplitude. Such form of  $\omega(b)$ can  also be related to  a   picture  of a convolution of the two energy--independent hadron pionic-type matter distributions (cf. Fig. 1, it illustrates  notion of centrality  in hadron scattering) in transverse plane:
 \begin{equation}
 \omega (b)\sim D_1\otimes D_2\equiv \int d {\bf b}_1 D_1({\bf b}_1)D_2({\bf b}-{\bf b}_1).
 \end{equation} 
 \begin{figure}[hbt]
 	\vspace{0cm}
 	\hspace{-1cm}
 	%	\begin{center}
 	\resizebox{16cm}{!}{\includegraphics{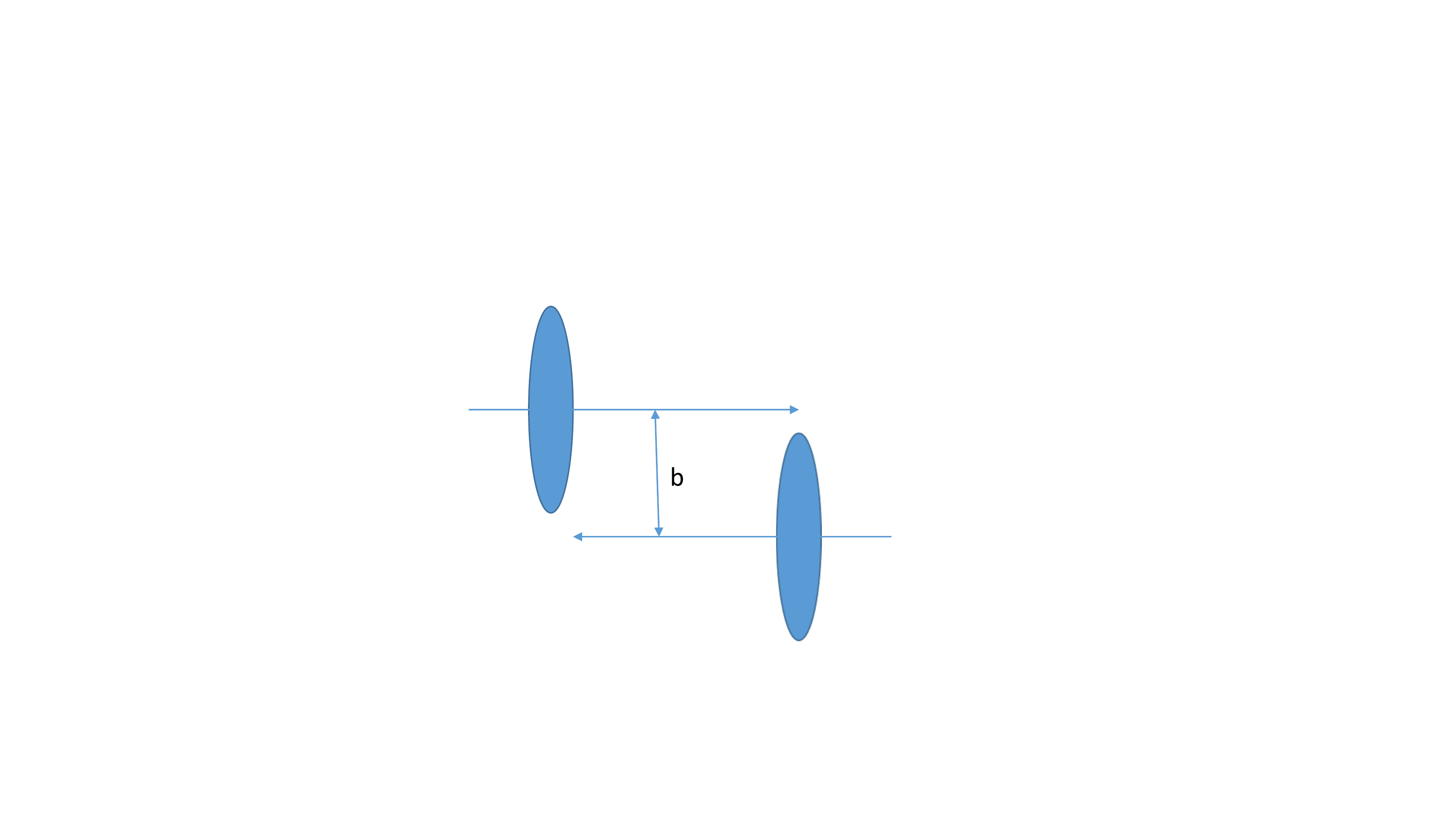}}		
 	%\includegraphics{dsdb2n.eps}
 	%	\end{center}
 	\vspace{-2cm}
 	\caption{Schematic view of  hadron scattering  with the  impact parameter $b$ in the geometric models (cf. e. g.  \cite{hei, chy, low,  sof}) .}	
 \end{figure}

 A  weak energy dependence of  centrality  allows one to use it as a parameter under the data analysis at  not too different energies in the hadron reactions.
 Indeed,  asymptotically, the centrality $c_b^h(s,b)$, defined according to Eq. (\ref{centhb}), decreases with energy slowly, like $1/\ln^2(s)$ at fixed impact parameter values. Unitarization leads to its dependence 
 $\sim b^2/\ln^2(s)$ since $f(s,b)$ saturates unitarity limit, i.e. $f(s,b)\to 1$ , at $s\to\infty$. Such slow energy decrease of centrality  allows one to compare the data at different energies and  approximately the same value of centrality provided     the energy values are high enough and  not too much different. Contrary, a strong energy dependence of centrality ($\sim 1/s^\lambda \ln s$ at $s\to\infty$ and fixed values of $b$) takes place  if it is defined through Eq. (\ref{centb}). It would make difficult a comparison of the data obtained at the same values of centrality  and even not too different energy values. 
 
 Now we would like to comment on  the experimental possibility of centrality measurements in hadron collisions. For that purpose
 using transverse energy deposited in a calorimeter seems to be a rather universal method since it includes the  case  of  unitarity saturation, i.e. it is supposed to determine centrality $c^h$ as a ratio of the transverse energy of the particular event $E_T^i$ to the total transverse energy of all events including elastic ones $\sum_i E_T^i$ , i.e.
 \[
 E_T^i/\sum_i E_T^i.
 \]
 
 ATLAS and CMS  experiments at the LHC are using transverse energy measurements in particular events  for the centrality determination in collisions of nuclei. Experimental feasibility of the similar measurements  for hadrons seems to be real.  Under the discussion of centrality measurements in hadron collisions one should keep in mind a  modification of centrality definition related to the essential role of elastic scattering component and the emergent black ring in  the inelastic collisions' probability distribution over impact parameter.

 \section*{Conclusion}

 The $b$-dependent differential quantities such as centrality are more sensitive to the  geometry of hadrons and their interaction dynamics than the quantities overintegrated over impact parameter. Centrality extracted from the experimental data can be used for the elastic scattering amplitude (its imaginary part) reconstruction in the impact parameter space according to Eq. (\ref{centhbder}).
 This relation of centrality with a $b$-space elastic amplitude   is  similar to the relation given by the optical theorem (it relates the elastic scattering amplitude with  properties of all the collisions including elastic and inelastic ones). 
 
   We conclude with notion that centrality is sensitive to a geometry of hadron interactions and an energy--dependent hadron interaction dynamics as well.

\section*{Acknowledgements}
We are grateful to E. Martynov for  the interesting discussions.


\small
\begin{thebibliography}{99}
\bibitem{rafel}
P. Koch, B. M\"uller, J. Rafelski,	Int. J. Mod. Phys. A {\bf 32}, 1730024 (2017) .
\bibitem{cms}
V.  Khachatryan et al. (The CMS Collaboration), Phys. Lett. B {\bf 765}, 193 (2017).
\bibitem{mangano}
B. Nachman, M. L. Mangano, Eur. Phys. J. C {\bf 78}, 343 (2018).
\bibitem{trib}
S. Schlichting, P. Tribedy,  Adv. High Energy Phys. {\bf 2016}, 8460349 (2016).
\bibitem{weili}
Wei Li, Mod. Phys. Lett. A {\bf 27}, 1230018 (2012).
\bibitem{weili1}
M. Floris, Wei Li,
 Adv. Ser. Direct. High Energy Phys. {\bf 29},  313 (2018). 
%\bibitem{jpg}
%S.M. Troshin, N.E. Tyurin, J. Phys. G: Nucl. Part. Phys. https://doi.org/10.1088/1361-6471/ab0ed1, in press.
\bibitem{olli}
S.J. Das, G. Giacalone, P.-A. Monard, J.-Y. Ollitrault, Phys. Rev. C {\bf 97}, 014905 (2018).
\bibitem{olli1}
R. Rogly, G. Giacalone,  J.-Y. Ollitrault, Phys. Rev. C {\bf 98}, 024902 (2018).
\bibitem{bron}
W. Broniowski and W. Florkowski, Phys. Rev. C {\bf 65}, 024905 (2002).
\bibitem{alkin1}
A. Alkin, E. Martynov, O. Kovalenko, S.M. Troshin, arXiv: 1807.06471v2.
\bibitem{umat}
V.I. Savrin, N.E. Tyurin, O.A. Khrustalev, Fiz. Elem. Chast. Atom. Yadra. {\bf 7}, 21 (1976).
\bibitem{2008}
S.M. Troshin, N.E. Tyurin, Mod. Phys. Lett. A {\bf 23}, 169 (2008).
\bibitem{rys}
V.A. Khoze, A.D. Martin, M.G. Ryskin, Phys. Lett. B {\bf 780}, 252 (1918).
\bibitem{gas}
R. Gastmans, S.L. Wu, T.T. Wu, Phys. Lett. B {\bf 720}, 205 (1913).
\bibitem{srvalue}
S.M. Troshin, N.E. Tyurin, Phys. Lett. B {\bf  316},  175 (1993).
\bibitem{factor}
S.M. Troshin, N.E. Tyurin, Mod. Phys. Lett. A {\bf 32}, 1750168 (2017).
\bibitem{hei}
W. Heisenberg, Zeit. Phys. {\bf 133}, 65 (1952).
\bibitem{chy}
T.T. Chou, C.N. Yang, Phys. Rev. Lett. {\bf 20}, 1213 (1968).
\bibitem{low}
F.E. Low, Phys. Rev. D {\bf 12}, 163 (1975). 
\bibitem{sof}
J. Soffer, C. Bourelly, T.T. Wu, AIP Conf. Proc. {\bf 1654}, 05008 (2015).





\end{thebibliography}
\end{document}